\documentclass[acmsmall,screen,nonacm]{acmart}
\usepackage{amsmath,amsthm,array}
\setcopyright{none}
\acmDOI{}
\acmISBN{}
\copyrightyear{}
\hypersetup{pdftitle={When Exact Inference Changes Code Selection: A Decision-Theoretic Analysis of B4},pdfauthor={YanliangCai}}
\newtheorem{theorem}{Theorem}
\newtheorem{proposition}[theorem]{Proposition}
\newtheorem{corollary}[theorem]{Corollary}

\theoremstyle{remark}
\newtheorem{remark}[theorem]{Remark}
\newcommand{\B}{\mathrm{B}}
\newcommand{\ind}{\mathbf{1}}

\newcommand{\Prob}{\mathbb{P}}

\title[When Exact Inference Changes Code Selection]{When Exact Inference Changes Code Selection: A Decision-Theoretic Analysis of B4}
\author{YanliangCai}
\affiliation{\institution{Independent Researcher}\country{China}}
\begin{document}
\begin{abstract}
When generated programs are ranked using generated tests, a more exact
posterior computation need not improve the selected program. We characterize
this distinction in the published B4 code--test model. Identical execution
signatures reduce exact candidate marginalization to $N$ summands and nonempty
joint maximum a posteriori (MAP) inference to two weights per group. A
prior--budget condition guarantees that B4's whole-group rule already attains
MAP; a sharp finite correction bound certifies when candidate marginals must
select the same group. All 2,421 default-setting comparisons on 807 development
matrices select the same candidates. A prospectively specified validation on
1,345 execution snapshots from a separate public archive also finds no changes
in 4,035 default-setting comparisons. Across 80 literature-derived priors,
this new archive yields 85 changed top sets in 107,600 comparisons; only six
coincide with failure of whole-group MAP. Under finite-test proxy labels,
four changes increase and five decrease expected success. These descriptive
changes span seven snapshots; per-prior mean differences range from
$-0.074$ to $+0.045$ percentage points. The results do not recommend a tuned
prior or a new performance method. Exact
enumeration and independent high-precision checks support the formulas and
decision diagnostics. A supplementary fixed-test analysis separates asymptotic
failure from observational ambiguity. The contribution is a model-specific
account of when an inference intervention can change a software-testing
decision, and when its effect must be sought in the priors or evidence instead.
\end{abstract}
\maketitle

\section{Introduction}
A test-guided coding workflow faces two uncertain objects: generated programs
and generated tests. Given a fixed pass/fail matrix, a developer can refine
posterior computation, change the prior, or acquire more informative tests.
These interventions address different causes of failure. A useful analysis
should identify which decision an inference routine optimizes, and whether
solving it more exactly can change the program delivered to the user.

We examine B4~\cite{b4}, an existing Bayesian model that scores groups of
programs with identical execution signatures. Its practical rule assigns
correctness to one entire group. Does that restriction miss the
highest-posterior nonempty correctness configuration? Even when it attains
that joint maximum a posteriori (MAP) value, does it maximize the probability
that the \emph{one selected program} is correct? The first is an optimization
question; the second depends on the decision loss. Neither establishes
whether the observations distinguish correct from incorrect programs.

Our starting point is the gap between stronger internal agreement and
stronger evidence for a delivered answer. We give that concern a narrow,
falsifiable form: an exact analysis of one published code--test model, with
no change to its generators or observation assumptions. The distinction
between MAP and marginal decisions is standard. The new work is the
model-specific reduction, endpoint characterization and sharp correction
bound, together with a diagnostic account of their consequences.

\begin{enumerate}
\item \textbf{RQ1: What does exact inference require?} Candidate marginal
scores need $N$ summands in total; nonempty joint MAP needs only two weights
per signature group (Theorems~\ref{thm:marginal} and~\ref{thm:map}).
The condition $a\ge N+b-3$ guarantees whole-group MAP optimality for $N\ge2$.
\item \textbf{RQ2: When can the decision be certified unchanged?} For all
positive Beta priors, a sharp uniform bound on marginal corrections yields a
label-free observable-margin certificate (Theorem~\ref{thm:bound},
Corollary~\ref{cor:certificate}, Proposition~\ref{prop:sharpness}). Worked
examples separate a MAP-value discrepancy from a candidate-choice discrepancy.
\item \textbf{RQ3: Where do these distinctions occur in existing archives?}
Exact marginalization changes none of 2,421 default-setting comparisons, but
changes 67 of 64,560 comparisons in a static 80-setting prior grid. Only 11 of
the 67 coincide with a whole-group MAP miss. These are reused development
matrices, not an independent test set or 64,560 independent tasks
(Section~\ref{sec:checks}). A subsequent prospective analysis of 1,345
CoSPlay snapshots finds no changes in 4,035 default comparisons, but 85
changes across the same 80-setting grid (Section~\ref{sec:independent}).
The new source is kept separate; snapshot and repeated-prior counts are
not treated as independent task counts.
\end{enumerate}

The results support a decision diagnostic rather than a new state-of-the-art
selector. They explain the negative default-setting result and delimit its
scope: a general positive-prior formula does not imply general agreement of
decisions. A supplementary fixed-test analysis separates changing a posterior
objective from identifying correctness across observationally equivalent
worlds (Appendix~\ref{sec:asymptotic}). Its large-budget failure family is not
evidence of practical few-candidate failure.

\section{Related work and scope}

\paragraph{The B4 model and its decision objective.}
B4 supplies the generative assumptions and whole-group scoring rule, and its
Sections 4 and 5.1 already analyze selector-dependent scaling. Thus neither
Bayesian test aggregation nor the existence of different scaling behavior is
our contribution. Its online Appendix A.1--A.6~\cite{b4appendix} concerns prior
perturbation and MaxPass/CodeT events. Our selected-candidate risk explicitly
includes tie handling, and differs from strict separation events used there.
The whole-group restriction is explicit in B4 Section 3.2: symmetry motivates
giving all members of a signature group the same predicted label. Symmetry of
marginal probabilities does not itself require a MAP configuration to label
every group member identically. Our endpoint result identifies when the
restriction preserves the MAP value, without treating it as an implementation
error or changing the original generative model.

\paragraph{Changing the estimator or the observations.}
Menet et al.~\cite{menet} study functional-similarity
estimators and feedback-driven generation, and include B4 comparisons.
In particular, their Section~3.5 relates code selection to correctness
probabilities, while Appendix~C.10 compares B4 with a soft MaxPass estimator.
Thus neither choosing by correctness probability nor contrasting B4 with
another selector is new here. Our narrower question keeps B4's likelihood
and priors fixed: can its own exact candidate marginal choose differently
from its whole-group score, and can agreement be certified from that matrix?
ExeCRE~\cite{execre} estimates reliability with a Dawid--Skene model over
execution-consistency signals. ExPairT-LLM~\cite{expairt} acquires new pairwise
membership and equivalence evidence. Those are broader or different inference
and observation mechanisms; this paper concerns exact calculations for a
fixed B4 matrix. We make no performance comparison to their complete systems.

\paragraph{Calibrating priors and evolving code--test populations.}
ZeroCoder~\cite{zerocoder} jointly trains code and test generators from
execution feedback. Its optional Dynamic B4 (DyB4) selector uses a small labeled
set to recalibrate B4's prior hyperparameters as training progresses. This
changes the prior and the generated matrices, whereas our archive comparison
keeps each matrix fixed and evaluates specified priors. DyB4's search grid includes candidate-prior strengths that
need not satisfy our sufficient MAP condition. The finite formulas apply
pointwise to a specified B4 matrix and positive hyperparameters, but neither
the default-archive agreement nor the fixed-prior asymptotics establish an
end-to-end result for adaptive calibration.

BACE~\cite{bace} evolves code and test populations using reciprocal Bayesian
belief updates anchored by immutable public input/output examples. Its noisy
sensor model permits a correct program to pass an incorrect test with positive
probability. That differs from B4's deterministic correct-row constraint, on
which our signature reduction depends. Its anchoring and evolution mechanisms
therefore concern a different inference and observation process. Our results
characterize exact decisions in B4 rather than the effectiveness of these
co-evolution and anchoring mechanisms. Label-free training, such as
ZeroCoder's static-selector setting, still uses execution observations; it
should not be equated with inference on an unchanging matrix.

CoSPlay~\cite{cosplay} co-evolves code and unit tests at test time using
bidirectional execution signals, and uses output consensus to break ties in
pass counts. We use its public final execution snapshots as a separate source
of matrices for our fixed-input diagnostic. This does not replace its full
selection procedure, replay its adaptive process, or establish that its
generated observations follow B4's likelihood.

\paragraph{Standard inference tools.}
The distinction between configuration MAP and marginal decisions, conjugate
integration, convexity of log moment-generating functions, score-margin
certificates, and two-world lower bounds are standard. Inconsistent hard-label
estimation is also classical; McLachlan and Ahfock~\cite{mclachlan} discuss it
for classification likelihood in mixture models. Our scope is the explicit
formulas and guarantees for this particular discrete selector, not a new
general theory of latent inference. The numerical findings are negative for
the proposed marginal correction as a practical method.

\begin{table}[t]
\caption{Closest mechanisms and the boundary of the present contribution.
This compares research questions, not system performance.}
\label{tab:scope}
\centering
\begin{tabular}{>{\raggedright\arraybackslash}p{0.15\linewidth}>{\raggedright\arraybackslash}p{0.34\linewidth}>{\raggedright\arraybackslash}p{0.40\linewidth}}
\toprule
Work & What changes or is analyzed & Relation to this paper\\
\midrule
B4~\cite{b4} & Bayesian model; whole-group rule; scaling & Inherited model and baseline; additional exact finite characterization.\\
Menet et al.~\cite{menet} & Soft similarity; correctness estimators; feedback limits & Prior decision-theoretic framing; our result fixes B4's model and bounds its marginal correction.\\
DyB4~\cite{zerocoder} & Prior calibration during generator training & Static diagnosis of a fixed snapshot; no training replication.\\
ExeCRE~\cite{execre} & Reliability from execution consistency & Different latent model; no complete-system comparison.\\
ExPairT~\cite{expairt} & New pairwise query evidence & Changes observations; our finite inference fixes the matrix.\\
BACE~\cite{bace} & Anchored co-evolution; noisy likelihood & Different likelihood and feedback; reduction does not transfer automatically.\\
\bottomrule
\end{tabular}
\end{table}

\section{Model, observations, and decision targets}
Let $N\ge1$ and $M\ge0$. The observed matrix $E\in\{0,1\}^{N\times M}$ records
whether candidate $i$ passes test $j$. Correctness variables are $X_i$ for
candidates and $Y_j$ for tests. We use the following hierarchical version of
the existing model. Independently draw
\[
\pi\sim\mathrm{Beta}(a,b),\quad
\eta\sim\mathrm{Beta}(a_y,b_y),\quad
\theta_t\sim\mathrm{Beta}(a_t,b_t),\ t\in\{0,1\},
\]
with all parameters strictly positive. Conditional on them, draw the $X_i$
iid Bernoulli$(\pi)$ and the $Y_j$ iid Bernoulli$(\eta)$. If $X_i=1$, row $i$
equals $Y$ deterministically. If $X_i=0$, its cells are independent given the
parameters and $Y$, with pass probability $\theta_{Y_j}$. These assumptions
are not asserted to hold for the real archives.

Write $C_N=\{\sum_iX_i>0\}$. We distinguish the unconditioned hierarchical
model from its entire joint law conditioned on $C_N$. This conditioning does
not preserve the original marginal prior on $\pi$. In particular, it differs
from drawing $\pi$ from that same prior and repeatedly sampling $X$ at that
fixed $\pi$ until $C_N$ holds. The latter inserts the factor
$[1-(1-\pi)^N]^{-1}$ inside the configuration integral; our four-Beta formulas
are not asserted for it.

The \emph{single-candidate} action selects an index $i$ with loss $1-X_i$.
The \emph{joint-MAP} target maximizes the posterior of a complete $(X,Y)$
configuration among nonempty consistent configurations. These losses need
not have the same optimal action.

Partition the rows into observed signatures $r\in\mathcal R$ of size $k_r$;
let $K=|\mathcal R|$ and $m_r=\sum_jr_j$. Define
\[
T_{1r}=\sum_{i,j:r_j=1}E_{ij},\qquad
T_{0r}=\sum_{i,j:r_j=0}E_{ij}.
\]
If a nonempty set of correct candidates has size $\ell$ and lies in group $r$,
consistency forces $Y=r$. After integrating the parameters, the weight of
\emph{one particular} such configuration, up to common positive constants, is
\begin{align}
w_r(\ell)={}&\B(a+\ell,N-\ell+b)\B(m_r+a_y,M-m_r+b_y)\nonumber\\
&\cdot\B(T_{1r}-\ell m_r+a_1,Nm_r-T_{1r}+b_1)\nonumber\\
&\cdot\B(T_{0r}+a_0,(N-\ell)(M-m_r)-T_{0r}+b_0).
\label{eq:weight}
\end{align}
All arguments are positive for $1\le\ell\le k_r$. The whole-group rule uses
$w_r(k_r)$; we write $S_r=\log w_r(k_r)$.

\subsection{A worked distinction between value and choice}
Consider four candidates with execution rows $00,01,01,11$ and all eight
Beta shapes equal to one. Table~\ref{tab:example} gives exact unnormalized
weights from \eqref{eq:weight}. The joint-MAP value is $1/6300$, attained
by either singleton group. The marginal numerator for \emph{each} candidate
in group $01$ is $1/17280+1/6480=11/51840$, so the optimal single-candidate
action instead chooses one of those two candidates. The nonempty normalizer
is $Z_+=533/907200$. The posterior success difference is $97/1066$,
not a measured accuracy improvement on programs. Group mass is not the same
as individual candidate correctness probability.

\begin{table}[t]
\caption{Exact constructed example. $A_r$ is per candidate, not per group.}
\label{tab:example}
\centering
\begin{tabular}{ccccc}
\toprule
Signature & Group size & $w_r(1)$ & $w_r(k_r)$ & $A_r$\\
\midrule
$00$ & 1 & $1/6300$ & $1/6300$ & $1/6300$\\
$01$ & 2 & $1/17280$ & $1/6480$ & $11/51840$\\
$11$ & 1 & $1/6300$ & $1/6300$ & $1/6300$\\
\bottomrule
\end{tabular}
\end{table}

Conversely, take rows $0,0$ and prior tuple
$(a,b,a_y,b_y,a_1,b_1,a_0,b_0)=(1,20,2,3,4,2,3,5)$.
The singleton weight is $1/31046400$ and the whole-group weight is
$1/194040000$, a ratio of $4/25$. The whole-group rule misses the joint-MAP
value, yet there is only one observed group. Its two candidates are
exchangeable, and exact marginalization cannot change their top set. A
MAP-value error alone therefore does not establish a selection error.

\section{Exact inference and the role of the whole group}
\begin{theorem}[Candidate marginal]
\label{thm:marginal}
For a particular candidate in group $r$, an unnormalized correctness
probability is
\begin{equation}
A_r=\sum_{\ell=1}^{k_r}\binom{k_r-1}{\ell-1}w_r(\ell).
\label{eq:marginal}
\end{equation}
Maximizing $A_r$ is the Bayes action for selecting one candidate, both in the
unconditioned model and after conditioning its entire joint law on $C_N$.
After sufficient-statistic construction, all marginals require $N$ summands.
\end{theorem}
\begin{proof}
If the selected candidate is correct, all other correct candidates must be in
its group and $Y=r$. With $\ell$ correct members, choose the other $\ell-1$
from $k_r-1$ possibilities. Every consistent state containing the candidate
appears once in \eqref{eq:marginal}. The nonempty-state normalizer is
\[
Z_+=\sum_{r\in\mathcal R}\sum_{\ell=1}^{k_r}
\binom{k_r}{\ell}w_r(\ell).
\]
Under conditioning on $C_N$, correctness probability is $A_r/Z_+$. Allowing
the all-incorrect states adds a common denominator term and changes no
ordering. Expected loss is minimized by the largest correctness probability.
Finally, $\sum_rk_r=N$. Column totals and signatures take $O(NM)$ operations,
and the group statistics take $O(KM)$, so the total is $O(NM+N)$ arithmetic
operations if a Beta evaluation is treated as a fixed-precision primitive.
This is not a bit-complexity or numerical-stability guarantee.
\end{proof}

\begin{theorem}[Two endpoints suffice for nonempty joint MAP]
\label{thm:map}
The unrestricted nonempty joint-MAP value is proportional to
\begin{equation}
\max_{r\in\mathcal R}\max\{w_r(1),w_r(k_r)\}.
\label{eq:map}
\end{equation}
For $N\ge2$, if $a\ge N+b-3$, the whole-group rule attains this value.
For $N=1$ the agreement is automatic.
\end{theorem}
\begin{proof}
Configurations within a group with the same $\ell$ have equal weight. Joint
MAP maximizes this individual weight; it does not multiply by the number of
such configurations. To locate its maximum, extend $w_r(\ell)$ to real
$\ell\in[1,k_r]$ through its integrated likelihood. Relative to the
$\ell=1$ integrand, the varying factor is $\exp((\ell-1)Z)$, where
\[
Z=\log\frac{\pi}{1-\pi}-m_r\log\theta_1
 -(M-m_r)\log(1-\theta_0).
\]
All integrals on this interval are finite: their Beta arguments interpolate
between positive endpoint arguments. H\"older's inequality implies
$w_r(t\ell+(1-t)u)\le w_r(\ell)^t w_r(u)^{1-t}$ for $0\le t\le1$.
Thus $\log w_r$ is convex and its maximum is attained at an endpoint, proving
\eqref{eq:map}.

For the sufficient condition, factor
$w_r(\ell)=\B(a+\ell,N-\ell+b)H_r(\ell)$.
Increasing $\ell$ by one removes success factors on columns where $r_j=1$
and failure factors where $r_j=0$ from the incorrect-candidate likelihood.
Each removed factor is at most one, so $H_r(\ell+1)\ge H_r(\ell)$.
The remaining ratio is
\[
\frac{\B(a+\ell+1,N-\ell-1+b)}{\B(a+\ell,N-\ell+b)}
 =\frac{a+\ell}{N-\ell+b-1}.
\]
It is at least one for every $1\le\ell\le N-1$ when
$a\ge N+b-3$. Each group's endpoint therefore attains its maximum.
\end{proof}

\begin{remark}
For $a=1000$, $b=1$ and $N\le100$, this sufficient condition holds regardless
of the other positive priors or the matrix. Consequently, an archive-level
comparison in that regime cannot attribute a difference between MAP and
marginal selection to a failure of the whole-group rule to attain MAP.
The candidate marginal remains a different objective. Outside the sufficient
condition, a singleton configuration can have higher posterior weight than
the full group; this need not change the selected group or its utility.
\end{remark}

\section{A finite correction bound and a selection certificate}
\begin{theorem}[General-prior correction bound]
\label{thm:bound}
Let $L_r=\log A_r$ and put $k=k_r$. For any positive Beta hyperparameters,
\begin{align}
0\le L_r-S_r&\le U_r
 :=\log\frac{\B(a+1,N-k+b)}{\B(a+k,N-k+b)}\label{eq:bound}\\
&\le\frac{(k-1)(N-k+b)}{a+1}
 \le\frac{(N+b-1)^2}{4(a+1)}.\nonumber
\end{align}
The bound is independent of hidden correctness labels.
\end{theorem}
\begin{proof}
The $\ell=k$ summand in \eqref{eq:marginal} yields $A_r\ge w_r(k)$.
The monotonicity of $H_r$ proved above gives
\[
\frac{A_r}{w_r(k)}\le
\frac{\sum_{\ell=1}^k\binom{k-1}{\ell-1}\B(a+\ell,N-\ell+b)}
 {\B(a+k,N-k+b)}.
\]
By the Beta integral and the binomial theorem, the numerator is
\begin{align*}
&\int_0^1 t^a(1-t)^{N-k+b-1}
  \sum_{\ell=1}^k\binom{k-1}{\ell-1}t^{\ell-1}(1-t)^{k-\ell}\,dt\\
&\hspace{2em}=\B(a+1,N-k+b).
\end{align*}
The resulting ratio is
$\prod_{j=0}^{k-2}[1+(N-k+b)/(a+1+j)]$, with empty product one.
Taking logarithms and applying $\log(1+x)\le x$ gives the first simpler
bound. The second follows from $t(D-t)\le D^2/4$ with
$t=k-1$ and $D=N+b-1$.
\end{proof}

\begin{corollary}[Unchanged selection]
\label{cor:certificate}
Suppose $r_*$ uniquely maximizes $S_r$. If
\begin{equation}
 S_{r_*}-S_r>U_r\quad\text{for every }r\ne r_*,\label{eq:certificate}
\end{equation}
then $r_*$ also uniquely maximizes $L_r$.
\end{corollary}
\begin{proof}
For each competitor, $L_r\le S_r+U_r<S_{r_*}\le L_{r_*}$.
\end{proof}

\begin{proposition}[Uniform sharpness of the Beta-ratio bound]
\label{prop:sharpness}
For any fixed matrix, group $r$, and positive candidate-prior parameters $a,b$,
the upper bound $U_r$ in \eqref{eq:bound} is the supremum of $L_r-S_r$ over
positive cell-likelihood prior parameters. This does not assert sharpness
for each fixed choice of those parameters.
\end{proposition}
\begin{proof}
Keep all other hyperparameters fixed and let $a_1=b_0=t\to\infty$.
Write $u=T_{1r}-k_rm_r+a_1$, $v=Nm_r-T_{1r}+b_1$,
$z=T_{0r}+a_0$, and $d=(N-k_r)(M-m_r)-T_{0r}+b_0$.
For $\ell\le k_r$, put $j_1=(k_r-\ell)m_r$ and
$j_0=(k_r-\ell)(M-m_r)$. With $(x)_j$ denoting the rising factorial,
\[
\frac{H_r(\ell)}{H_r(k_r)}
=\frac{(u)_{j_1}}{(u+v)_{j_1}}
 \frac{(d)_{j_0}}{(z+d)_{j_0}}\longrightarrow1.
\]
The finite candidate-prior sum then tends to the Beta ratio in the proof
of \eqref{eq:bound}. Taking logarithms proves the asserted supremum.
The case $k_r=1$ already has equality at every finite parameter value.
\end{proof}

The certificate compares each competing group's largest allowed correction
with its current deficit. Its failure does not predict a change or improvement.
Tied endpoint winners are not certified by this sufficient condition; a
one-group matrix is trivially unchanged. At $a=1000,b=1$, the universal bound
is approximately $0.02498$ for $N=10$ and $2.49751$ for $N=100$. These are
log-score bounds, not percentage-point accuracy guarantees.

\paragraph{How to use the certificate.}
For a fixed matrix and prior, compute the whole-group log-scores and each
competitor's bound $U_r$. If the winner is unique and all inequalities in
\eqref{eq:certificate} hold, exact marginalization cannot change the selected
group. If any inequality fails, compute \eqref{eq:marginal}; failure of a
sufficient certificate is not evidence of a changed decision. In particular,
this check cannot establish whether the common selected group is correct.
It isolates a possible computational intervention from changing the evidence
or prior. We claim no runtime advantage: the marginal sum is already tractable,
and computing the certificate need not be cheaper in a given implementation.

\paragraph{A reproducible decision diagnostic.}
For fixed priors: (1) group identical rows and record multiplicities;
(2) compare $w_r(1)$ and $w_r(k_r)$ to check whole-group MAP optimality;
(3) apply the margin certificate, separating tied and single-group cases;
(4) if it does not certify agreement, compute $A_r$ using log-sum-exp;
(5) compare the full top candidate sets. If utility is evaluated, sample
uniformly among top candidates, not groups. Hidden labels are used only for
that final evaluation. A failed certificate means ``compute the marginal'',
not ``the marginal improves correctness''. We claim no speedup.

\section{Computational checks and prior sensitivity}
\label{sec:checks}
\paragraph{Method and provenance.}
RQ1 is checked against nonempty latent-state enumeration; RQ2 against exact
rational bounds and observed margins; RQ3 compares full top sets with fixed
inputs and priors. The three-setting analysis preceded the correction bound,
whose coverage is explanatory. Before running the additional grid, we fixed
its parameters, input hashes, outcomes and refinement rules in a local
protocol. This prospectively specifies a new analysis of already-inspected
data; it is not public preregistration or held-out validation.

\paragraph{Use of generative AI in the research process.}
OpenAI Codex was used as an interactive LLM assistant to propose and audit derivations,
write numerical code, inspect public literature and revise text. It also
assisted with the manuscript revision, public-archive discovery, source and schema
inspection, acquisition/qualification code, frozen analysis protocols and
their documented resource/schema clarifications, numerical analysis and
artifact checks. No LLM was queried to
generate new benchmark candidates or tests for the reported experiments.
The evidence consists of explicit proofs, executable arithmetic checks and
archived observations, not model judgments of novelty or simulated reviewer
scores. The author is responsible for these contributions; computational
agreement is not independent peer review.

\paragraph{Exact finite checks.}
An initial enumeration covers all 209 two-column row multisets with
$1\le N\le6$ under four fixed priors. All 836 grouped-versus-latent comparisons
and 4,032 candidate numerators agree exactly using rational arithmetic.
There are no selection changes under the three default prior settings.
Only the all-one-prior control changes choices. Its smallest example has
rows $00,01,01,11$: whole-group scoring ties $00$ and $11$, whereas candidate
marginals favor $01$. Uniform selection among the top whole-group candidates
loses $97/1066$ in model-conditional posterior success. This is not observed
benchmark gain, and the control is not a recommended default prior.

A subsequent check enumerates all row multisets for $M=0,1,2$, $1\le N\le6$,
with eight positive integer prior tuples. The 1,936 matrix--prior panels and
4,416 groups confirm the marginal formula, two-endpoint MAP value, discrete
log-convexity, and general Beta-ratio bound against explicit latent-vector
enumeration. All 1,217 panels satisfying the sufficient MAP condition attain
MAP with whole groups. Outside that condition, 184 panels miss MAP; the first
nonempty-test witness has rows $0,0$ and prior tuple
$(a,b,a_y,b_y,a_1,b_1,a_0,b_0)=(1,20,2,3,4,2,3,5)$, where the full-group to
MAP weight ratio is $4/25$. These MAP misses need not imply selection losses.
The observational twin is separately checked on 4,092 ordered matrices and
19,116 actions using exact fractions, including the conditioning distinction.
Finite checks supplement the analytic proofs; they do not prove universality.

\paragraph{Archived inputs.}
We reuse 497 CodeMonkeys repository-task matrices~\cite{codemonkeys}, 155
nonempty-test CodeT function-task matrices~\cite{codet}, and the same 155
functions after retaining only tests accepted by a reference implementation.
The latter is an additional-oracle channel, not an independent dataset.
Its 14 empty-test matrices remain in the denominator. Repository matrices
have at most 10 candidates; function matrices have 100. These are previously
inspected development data, not a newly held-out test set. Candidate/test
columns and their multiplicities are frozen. No candidate generation or
model calls are performed in this analysis.

\paragraph{Selector comparison.}
We compare two scoring objectives on the same supplied matrix, not the full
published B4 experimental pipeline. At the pinned implementation revision,
\texttt{strategy.B4} scores all supplied signature groups. Its experiment
driver separately filters tasks by held-out correctness rate and excludes
unsuccessful execution records~\cite{b4implementation}. The execution
record's \texttt{passed} flag in that path denotes a nonempty result list,
not a positive test bit; it must not be read as a zero-pass-row filter.
Our frozen archives retain zero-pass rows, candidate multiplicities, empty
generated-test sets and tasks without a positive label. Consequently these
counts do not reproduce B4's published pass@1 denominator or preprocessing.

All three fixed settings use $a=a_y=1000$, $b=b_y=a_1=b_1=a_0=1$, and
$b_0\in\{10^4,10^5,10^6\}$. Hidden labels are not supplied to either selector.
Every top candidate set is unchanged after exact marginalization, hence
accuracy changes are exactly zero under both the existing benchmark labels
and an additional conjunctive EvalPlus~\cite{evalplus} labeling sensitivity check. The 807
matrices yield 2,421 comparisons, not 2,421 independent tasks.
Scores are calculated at 80 decimal digits; a separate implementation
expands repeated columns and recomputes choices at 120 digits without
importing the selector. All top sets agree.

\begin{table}[t]
\centering
\caption{Certificate coverage over the three prior settings. ``Margin''
counts multi-group comparisons satisfying \eqref{eq:certificate}. No selection
changes occur, including in the one uncertified comparison.}
\label{tab:certificate}
\begin{tabular}{lrrrr}
\toprule
Channel & Comparisons & One group & Margin & Uncertified\\
\midrule
Repository & 1,491 & 534 & 957 & 0\\
Function, raw & 465 & 27 & 438 & 0\\
Function, reference-compatible & 465 & 45 & 419 & 1\\
\midrule
Total & 2,421 & 606 & 1,814 & 1\\
\bottomrule
\end{tabular}
\end{table}

The finite bound certifies 1,814 multi-group cases, and 606 cases have only
one group (Table~\ref{tab:certificate}). All 16,020 observed group corrections
satisfy the bound numerically. The remaining comparison is HumanEval/105 in
the reference-compatible channel with $b_0=10^5$; direct marginal computation
also leaves it unchanged. The bound was derived after observing the zero-change
result, so this coverage is explanatory, not preregistered validation.
Coverage applies the analytic certificate numerically at high precision;
it is not an interval-arithmetic verification of the archive.
We stopped empirical expansion of marginalization as a performance method
instead of tuning priors to manufacture a positive result.

\subsection{A literature-derived static prior grid}
ZeroCoder v2 Section 3.3~\cite{zerocoder} gives DyB4's search ranges
$a=a_y\in\{10^0,\ldots,10^7\}$ and
$b_0\in\{10^0,\ldots,10^9\}$, with other shapes one. We evaluate all
80 settings on all 807 frozen matrices, including empty-test matrices.
This diagnoses the static prior range; it does not reproduce labeled
calibration, online training or changing candidate populations. No setting
is selected by its measured success.

The grid yields 64,560 matrix--prior comparisons. Whole-group scores miss
the nonempty MAP value in 71 comparisons, all with $a=1$. Marginalization
changes 67 top candidate sets across 42 matrices; only 11 changes coincide
with a MAP miss. In 16 changed comparisons the sufficient MAP condition
itself holds. These observations directly separate a change of decision
objective from correcting a MAP approximation. There are no top-set changes
for $a\ge1000$ anywhere in this grid, although this remains a finite-archive
observation, not a theorem about those priors.

\begin{table}[t]
\caption{Static prior grid on reused development matrices. Each prior value
includes 10 values of $b_0$. Up/down/equal describes archived-label success
among changed comparisons, using uniform selection among top candidates.}
\label{tab:grid}
\centering
\begin{tabular}{lrrrr}
\toprule
$a=a_y$ & Comparisons & MAP-value misses & Top-set changes & Up/down/equal\\
\midrule
$1$ & 8,070 & 71 & 40 & 4/11/25\\
$10$ & 8,070 & 0 & 23 & 3/7/13\\
$100$ & 8,070 & 0 & 4 & 2/1/1\\
$10^3,\ldots,10^7$ & 40,350 & 0 & 0 & 0/0/0\\
\midrule
Total & 64,560 & 71 & 67 & 9/19/39\\
\bottomrule
\end{tabular}
\end{table}

The sufficient prior--budget condition covers 53,390 comparisons. The margin
certificate covers 47,675 multi-group comparisons, and 16,160 have one group.
The remaining 725 include all 67 top-set changes; failure of the certificate
is not itself predictive of a change. Under archived labels, 9 changes
improve uniform-tie success, 19 reduce it and 39 leave it unchanged
(Table~\ref{tab:grid}). The counts describe repeated matrices and correlated
priors, not independent observations, a significance test or evidence for a
recommended performance intervention.

\paragraph{Numerical verification.}
Before applying the grid, the screening implementation was compared with
explicit latent-vector enumeration on all row multisets with $N\le6$ and
$M\le2$, including $M=0$, for all 80 priors: 19,360 matrix--prior comparisons.
The largest absolute log-marginal discrepancy was below $1.05\cdot10^{-8}$.
Archive screening uses double-precision log-Beta and log-sum-exp with a
conservative $10^{-4}$ refinement threshold. Every apparent decision change,
MAP miss, nontrivial near tie and near certificate boundary is recomputed
from literally expanded columns at 80 decimal digits. The 139 flagged
comparisons use a $10^{-55}$ decision tolerance. This is numerical checking,
not interval-arithmetic proof. All 139 are also rechecked at 120 digits;
240 additional weighted-column and empty-test comparisons agree with exact
latent enumeration. All reported changes pass refinement.

\paragraph{Archive discovery.}
The official B4 archive~\cite{b4appendix} supplies generated solutions and
tests but no separately listed executed matrices. We did not treat these
generated-source files as new execution evidence. A subsequent search found
the CoSPlay cache, evaluated separately below rather than pooled with the
already-inspected development matrices.

\subsection{Validation on an independent-source execution archive}
\label{sec:independent}
\paragraph{Protocol and provenance.}
After fixing the theory and development analysis, we identified the public
CoSPlay execution archive~\cite{cosplay}. Before downloading its selected
JSON records or computing B4 scores, we fixed all final snapshot shards for
the released Cosplay-7b model's run~0 across LiveBench, CodeContests,
CodeForces and LiveCodeBench: 29 files at Hub revision
\texttt{bac85aaa52282a097ab82d45464819cfffe62908}.
We did not select tasks by their labels or selector differences. The
80-setting grid is unchanged; the three original defaults are the primary
comparison. A metadata-driven resource amendment increased the download
cap while preserving all selected files. This is a local prospective
protocol, not a public preregistration.

Pinned upstream execution code binds each final candidate to the same row
in both generated-test and held-out-test matrices, and reruns executions
after candidate refinement. We checked rectangularity, candidate counts,
redundant row encodings, valid-test masks, and input/output column counts.
Only generated columns marked valid by the archive are retained; candidates,
repeated rows and repeated tests are preserved. In particular, zero-pass
rows are not discarded. Original file lengths and LFS SHA256 digests are
verified before records enter the analysis. No generated program is executed
by our analysis, and no generator is called.

\paragraph{Label and sampling limits.}
The cached held-out matrix can contain only a prefix of the benchmark's
available tests. Passing its nonempty columns defines a \emph{proxy} label,
not passing the complete benchmark or establishing semantic correctness.
An empty held-out matrix would have unknown labels, not automatic success.
Where the archive omits task identifiers, a benchmark-specific question hash
and the source file/row identify the record. Identical question hashes are
checked within and across benchmarks. Independence here means an additional
source not used to develop the theory; it does not mean independent human
replication, independent observations across priors, or a new execution of
the CoSPlay pipeline. We evaluate B4 on final snapshots, not CoSPlay's full
selection or adaptive process.

\paragraph{Qualified inputs and model diagnostic.}
All 29 shards pass source-hash and schema checks, yielding 1,345 snapshots
with 16 candidates each. We retain 18,354 valid generated-test columns and
exclude 3,166 columns marked invalid; seven zero-column snapshots remain.
No snapshot has unknown proxy labels, and 774 have no proxy-positive
candidate. The cached held-out matrix has between one and eight columns;
914 snapshots contain fewer executed columns than available benchmark tests.
There are 1,313 distinct benchmark--question-hash keys and 1,217 distinct
question hashes across benchmarks. In particular, 32 LiveBench keys occur
twice. All occurrences remain as specified, with duplicate mappings in the
artifact; these are snapshot-level descriptive counts, not independent
sampling units.

In 156 snapshots, at least two proxy-positive candidates have different
generated-test signatures (25 LiveBench, 18 CodeContests, 18 CodeForces and
95 LiveCodeBench). If these proxy labels were treated as true correctness,
they would violate B4's deterministic correct-row constraint. They are not
true semantic labels: incomplete held-out coverage, execution effects and
model misspecification are not distinguished by this check. The observation
limits the interpretation of model-based probabilities; it does not refute
the conditional mathematical results.

\paragraph{Primary and sensitivity results.}
All 4,035 default-setting comparisons preserve the full top candidate set.
Of these, 1,038 have one signature and 2,982 are certified by the margin
bound; direct computation verifies agreement in the remaining 15.
The prior--budget sufficient condition holds in every default comparison.
Thus the negative default result extends to these final co-evolution
snapshots, without establishing a universal equivalence.

\begin{table}[t]
\centering
\caption{Independent-source snapshots, all 80 frozen priors. Changes count
full top candidate sets. Up/down/equal concerns finite-test proxy-label
success among changed comparisons; neither tasks nor priors are independent.}
\label{tab:independent}
\begin{tabular}{lrrrrr}
\toprule
Archive & Snapshots & Comparisons & MAP misses & Changes & Up/down/equal\\
\midrule
LiveBench & 128 & 10,240 & 0 & 3 & 1/0/2\\
CodeContests & 239 & 19,120 & 25 & 13 & 1/1/11\\
CodeForces & 467 & 37,360 & 67 & 53 & 0/4/49\\
LiveCodeBench & 511 & 40,880 & 15 & 16 & 2/0/14\\
\midrule
Total & 1,345 & 107,600 & 107 & 85 & 4/5/76\\
\bottomrule
\end{tabular}
\end{table}

Across all 80 settings, marginalization changes 85 top sets on 60 snapshots;
only six coincide with a whole-group MAP-value miss. Five changes occur
despite the sufficient MAP condition. There are 55, 25 and five changes at
$a=a_y=1,10,100$, respectively, and none at $a\ge1000$. All 107 MAP misses
occur at $a=1$. Four changes improve proxy-label success under uniform
candidate tie handling, five reduce it and 76 leave it unchanged
(Table~\ref{tab:independent}). This is evidence that exactness and delivered
utility are distinct, not an accuracy advantage for the marginal selector.

The unchanged frozen numerical procedure refines 459 flagged comparisons
at 80 digits; all 459, including every change and MAP miss, agree when
rechecked at 120 digits. The 240 exact weighted-column/empty-test checks also
pass. Neither the data acquisition nor this analysis makes model calls or
reruns the archived candidate programs.

\paragraph{Effect magnitude and analysis units.}
A post-result descriptive reanalysis reconstructs candidate-level utility
for all 107,600 comparisons from the frozen top sets and labels; every
stored difference agrees exactly. Only nine comparisons, on seven snapshots,
have nonzero utility differences. For each fixed prior separately, let
$\Delta$ be mean marginal-selector success minus mean whole-group success.
Table~\ref{tab:effect-size} gives the range of these 80 means, not a confidence
interval, a tuned-prior estimate, or the performance of a mixture of priors.
All three default-prior differences are exactly zero under every weighting.
Repeated snapshots are averaged within each key before averaging keys; the
question-only sensitivity also groups matching hashes across benchmarks.
Neither weighting establishes statistical independence or removes unknown
semantic duplicates.

\begin{table}[t]
\centering
\caption{Post-result sensitivity of proxy utility. Ranges of per-prior
$\Delta$ across all 80 fixed priors, in percentage points (pp). The
positive-candidate subset uses labels only for this descriptive analysis.}
\label{tab:effect-size}
\begin{tabular}{lrrr}
\toprule
Weighting / population & Units & Minimum (pp) & Maximum (pp)\\
\midrule
Equal snapshot & 1,345 & $-0.07435$ & $+0.04461$\\
Equal benchmark--question & 1,313 & $-0.07616$ & $+0.04570$\\
Equal question hash & 1,217 & $-0.08217$ & $+0.04930$\\
Snapshots with a proxy-positive candidate & 571 & $-0.17513$ & $+0.10508$\\
\bottomrule
\end{tabular}
\end{table}

The 774 snapshots without any proxy-positive candidate cannot yield a
positive proxy outcome under any selector restricted to these candidates.
The conditional subset exposes this ceiling without changing the frozen
primary denominator. Small aggregate differences do not imply small changes
on each task: the nine nonzero differences range from $-1$ to $+3/5$ in
success probability. These results support a finite decision distinction,
but supply no stable accuracy advantage for exact marginalization. The
artifact records all per-prior means, including zero and negative results.

\paragraph{A real snapshot separating MAP and the delivered action.}
For illustration we take the first frozen panel--prior record with a changed
top set despite the sufficient MAP condition, without selecting on utility.
This post-result example is CodeContests task 4081, source chunk~2, row~25
(zero-based), with $N=16$, $M=15$, $a=a_y=100$, $b_0=100$, and the other
shape parameters one. Table~\ref{tab:real-witness} reports every signature;
all scores are recomputed at 120 decimal digits and rounded for display.
Because $100\ge16+1-3$, whole-group scoring attains the nonempty MAP value.
Yet a group of 12 candidates receives marginal correction $0.019394$,
exceeding its whole-group deficit $0.013730$, and becomes the selected group.
The certificate appropriately fails: its competitor bound is $0.507434$.
All 16 candidates fail at least one of the eight cached held-out tests
(ten tests are available upstream), so the action changes without a proxy
utility gain. This is a concrete decision discrepancy, not an improved
generated solution or a failure of the MAP optimization.

\begin{table}[t]
\centering
\caption{Complete execution signatures for the illustrative snapshot.
$S-S_{\max}$ and $L-L_{\max}$ are relative log-scores, with separate maxima.
No candidate is proxy-positive.}
\label{tab:real-witness}
\begin{tabular}{lrrr}
\toprule
Signature & Size & $S-S_{\max}$ & $L-L_{\max}$\\
\midrule
\texttt{000000000000000} & 1 & $-79.990494$ & $-79.996158$\\
\texttt{100000000000000} & 12 & $-0.013730$ & $0.000000$\\
\texttt{100010100000000} & 1 & $-58.836635$ & $-58.842298$\\
\texttt{111111011111111} & 1 & $-9.598817$ & $-9.604480$\\
\texttt{111111111111111} & 1 & $0.000000$ & $-0.005664$\\
\bottomrule
\end{tabular}
\end{table}

\section{What the characterization does and does not resolve}
The finite results supply a concrete distinction between three questions:
whether a rule attains joint MAP, whether that MAP-based rule agrees with
the single-candidate Bayes action, and whether the observations identify the
correct answer across possible worlds. In the studied default-budget regime,
the first is guaranteed, the second is observed and almost entirely certified,
and the third is not guaranteed by the model. This explains why a formally
more appropriate marginal objective can be computationally feasible yet
empirically inactive.

The limit theorem concerns a fixed observation alphabet and fixed prior
strength. It supplies no practical convergence rate, no result for growing
test counts, and no winner claim at tied limiting scores. Actual code/test
generation may violate conditional independence, share systematic errors,
or change with feedback. None of those mechanisms is identified by this
analysis. Source-code inspection, verified test labels, and adaptive tests
can alter the observation channel and are outside the two-world bound.
Likewise, DyB4-style recalibration can alter the prior. Agreement at the three
default settings does not establish agreement under its selected parameters.
Our static grid shows that choices can change within its search range, but
does not predict which priors its adaptive procedure selects or its
end-to-end accuracy. No adaptive-system experiment is reported here.
The result therefore cannot establish that additional internal computation
is useless or that every collaboration system requires a particular external
oracle. It addresses the candidate-selection component alone.

The contribution is an explicit, reproducible theoretical characterization
of an existing model, with a negative default-setting result that also holds
on an additional source specified before its outcomes were read, and a finite
prior-sensitivity diagnostic. It is not a new
state-of-the-art selector. A practical intervention would need a separately
specified observation or model change and a new controlled evaluation;
the current evidence does not justify that claim.

\subsection{Threats to validity and practical use}
\textbf{Construct validity.} A test pass is an observed bit, not correctness.
B4 makes a deterministic claim about correct rows, and shared generation
errors can violate its likelihood. Exact inference under a misspecified
model is not an assurance of program correctness. Candidate success uses
archived labels; reference-compatible filtering adds oracle information
and is not independent of the raw channel.

\textbf{Internal validity.} The default-setting result motivated the
correction bound. The later frozen grid tests a broader finite domain but
cannot remove that development history. Independent counting, exact finite
checks and precision refinement reduce implementation risk; they cannot
replace proofs or external mathematical review. Empty test sets, group
multiplicities and ties are preserved without outcome-dependent filtering.
The effect-size, task-weighting and real-snapshot analyses were added after
the grid outcomes were known and are explicitly exploratory. They do not
retroactively extend the prospective validation protocol.

\textbf{External validity.} The 807 development matrices were previously
examined; the later CoSPlay source was specified before its outcomes were
read. This broadens the snapshots, not the number of independent research
teams or model runs. Duplicate questions, correlated candidates and repeated
priors preclude treating comparison counts as independent trials. Limited
held-out tests define proxy labels and reveal potential mismatch with the
correct-row assumption. The study does not rerun generators, adaptive testing
or online prior selection. The asymptotic construction fixes the number of tests; its
examined transitions are far beyond practical candidate budgets. The
diagnostic is appropriate when an engineer already uses a B4 snapshot:
it can establish whether inference changes a choice, while better delivered
software still needs an appropriate evaluation.

\section*{Data Availability}
A local reproducibility package contains manuscript source,
frozen matrix inputs, exact finite checkers, prior-grid code, numerical
results and a hash manifest. The matrix sources are CodeMonkeys,
CodeT and CoSPlay~\cite{codemonkeys,codet,cosplay}; reference-compatible
filtering is separately identified. The CoSPlay portion includes compact
execution extracts, source receipts, fixed revisions, qualification reports,
duplicate mappings, and a streaming acquisition script; full generated code
and execution logs are not redistributed. Upstream attribution and license
notices are retained. The new analysis makes no model calls and executes no candidate
programs. No public artifact URL has yet been assigned. The analysis code and compact execution extracts have not yet
been publicly released; public availability of the upstream sources is not
equivalent to public availability of our processed analysis package. The generated-code-only B4 archive is cited for
provenance and is not represented as a new executed dataset.

\appendix
\section{Fixed-test asymptotics and observational ambiguity}
\label{sec:asymptotic}
The previous results are finite algebraic statements. We now hold $M$ and all
positive hyperparameters fixed as $N$ grows. This regime excludes increasing
test counts and priors whose strength grows with $N$.

Let $q_r$ be the empirical row frequency. For an observed signature $r$, set
\begin{align*}
d_{1r}(q)&=\sum_s q_s\sum_{j:r_j=1}s_j-q_rm_r,&
d_{0r}(q)&=\sum_s q_s\sum_{j:r_j=0}s_j,\\
v_{1r}(q)&=(1-q_r)m_r,&v_{0r}(q)&=(1-q_r)(M-m_r).
\end{align*}
Define $\Phi(u,v)=u\log(u/v)+(v-u)\log((v-u)/v)$ on $0\le u\le v$,
using continuous endpoint values, including $\Phi(0,0)=0$.

\begin{theorem}[Normalized whole-group score]
Uniformly over empirical distributions $q$ and observed signatures,
\begin{equation}
\frac{S_r}{N}=F_r(q)+O\!\left(\frac{\log(N+2)}N\right),\qquad
F_r(q)=\Phi(q_r,1)+\Phi(d_{1r},v_{1r})+\Phi(d_{0r},v_{0r}).
\label{eq:limit}
\end{equation}
If rows are iid with law $p$ and $F_r(p)$ has a unique maximizer $r_*$ on
the positive support of $p$, B4 eventually selects $r_*$ almost surely.
\end{theorem}
\begin{proof}
For each fixed $c>0$, log-Gamma asymptotics~\cite{nist} and a bounded
initial-range argument give
$\log\Gamma(t+c)=t\log t-t+O(\log(t+2))$ uniformly for integers $t\ge0$,
with $0\log0=0$. For fixed positive $c,d$ and integer $u,v\ge0$ bounded
by $\max(1,M)N$, it follows that
\[
\log\B(u+c,v+d)=\Phi(u,u+v)+O(\log(N+2)).
\]
The constants may depend on the fixed hyperparameters and $M$, but not on
the counts. Apply this to the candidate-prior and two cell-likelihood factors
in \eqref{eq:weight}, with $\ell=k_r$. The test-prior factor has bounded
counts and contributes $O(1)$. This proves \eqref{eq:limit}, including empty
blocks. There are finitely many signatures. The strong law gives $q\to p$
almost surely, and $\Phi$ is continuous on its compact domain. Every supported
signature eventually appears. The strictly positive limiting score gap then
exceeds the uniform vanishing remainder, proving eventual selection.
\end{proof}

Fix actual parameters and a test vector $y$ for the following risk calculation.
Writing $g$ for the incorrect-row product Bernoulli law,
$p_r=\pi\ind\{r=y\}+(1-\pi)g_r$.
Conditional on the observed matrix, a candidate in row $y$ is correct with
probability $\pi/p_y$; candidates in other rows are incorrect. Independence
of other rows at fixed parameters and the law of total expectation imply,
for any matrix-only selector $\delta$ with independent randomization,
\begin{equation}
\Prob(X_{\delta(E)}=1)=\frac{\pi}{p_y}
 \Prob(E_{\delta(E),:}=y).\label{eq:risk}
\end{equation}
The selector need not know those parameters. Equation~\eqref{eq:risk} is a
fixed-parameter risk identity, not a prior-predictive Bayes-risk assertion.

\begin{proposition}[A conditional failure family]
Let $M=2$, $y=(1,1)$, $\theta_1=1/2$, and $\pi=1-4\varepsilon$ for
$0<\varepsilon\le1/100$. For every fixed positive B4 prior, its
selected-candidate success tends to zero. MaxPass, which selects a row with
the most ones, has limiting success
$c=(1-4\varepsilon)/(1-3\varepsilon)$.
\end{proposition}
\begin{proof}
The row probabilities in order $00,01,10,11$ are
$(\varepsilon,\varepsilon,\varepsilon,1-3\varepsilon)$. With binary entropy
$h(z)=-z\log z-(1-z)\log(1-z)$, substitution into \eqref{eq:limit} gives
\begin{align*}
F_{00}&=-h(\varepsilon)-2(1-\varepsilon)h(\varepsilon/(1-\varepsilon)),\\
F_{11}&=-h(3\varepsilon)-6\varepsilon h(1/3),\\
F_{01}=F_{10}&=-h(\varepsilon)-(1-\varepsilon)h(\varepsilon/(1-\varepsilon))\\
&\hspace{4em}-(1-\varepsilon)h(2\varepsilon/(1-\varepsilon)).
\end{align*}
Since $h$ increases strictly below $1/2$, $F_{00}>F_{01}=F_{10}$.
For $G=F_{00}-F_{11}$, $G(0)=0$ by continuous extension, and differentiation
yields
\[
G'(\varepsilon)=\log\left[\frac{27}{16}
 \frac{(1-3\varepsilon)^3(1-\varepsilon)}{(1-2\varepsilon)^4}\right]>0.
\]
Indeed, the argument is at least
$(27/16)(97/100)^3(99/100)>1$ throughout the interval. Thus $00$ uniquely
maximizes the limit and is incorrect. Apply \eqref{eq:risk}. MaxPass selects
$11$ whenever present, whose probability tends to one, yielding success $c$.
\end{proof}

\paragraph{Finite-budget boundary.}
At $\varepsilon=1/4004$, $\pi=1000/1001$ equals the default candidate-prior
mean. Fixing means is not sampling from the hierarchical prior. In a
deterministic population-count grid $N=4004\cdot10^j$, the first observed
change to the wrong winner occurs at $N=4{,}004{,}000{,}000$ for
$b_0=10^4$, ten times later for $10^5$, and another ten times later for
$10^6$, with $a=a_y=1000$ and other shape parameters one. These are grid
points, not exact transition thresholds or executed candidate populations.
They provide no evidence of practical few-agent failure.

\begin{proposition}[An observational twin]
For the family above, a second world with $y=(0,0)$,
\[
\pi'=\frac{\varepsilon(1-4\varepsilon)}{1-3\varepsilon},\qquad
\theta'_0=\frac{1-3\varepsilon}{1-2\varepsilon}
\]
has the same unconditioned matrix law for every $N$, but the opposite correct
signature. For any matrix-only selector, its successes in the two worlds sum
to at most $c$. The asymptotic minimax success over these two worlds is $c/2$.
\end{proposition}
\begin{proof}
Put $t=\theta'_0$. Since
$1-\pi'=(1-2\varepsilon)^2/(1-3\varepsilon)$ and
$1-t=\varepsilon/(1-2\varepsilon)$, the second world's probabilities are
\[
p'_{11}=(1-\pi')t^2=1-3\varepsilon,\quad
p'_{01}=p'_{10}=(1-\pi')t(1-t)=\varepsilon,\quad
p'_{00}=\pi'+(1-\pi')(1-t)^2=\varepsilon.
\]
Rows are iid, proving equality of matrix laws. In each world, a member of
its correct signature is correct with conditional probability $c$. Under
the common observation law let $q_{11}$ and $q_{00}$ be the probabilities
that the selector chooses these signatures. Its two successes are $cq_{11}$
and $cq_{00}$, whose sum is at most $c$. Randomization is included; no
permutation-equivariance assumption is needed. Both signatures eventually
appear with probability tending to one. Flipping a fair coin between them
when both appear, and choosing any available candidate otherwise, attains
success tending to $c/2$ in each world.
\end{proof}

\paragraph{Conditioning and interpretation.}
Conditioning each fixed world on $C_N$ generally destroys exact finite-$N$
observational equality. With $a_W=1-(1-\pi_W)^N$, success implies $C_N$,
so the valid conditional bound is
\[
a_A s_{A\mid C_N}+a_B s_{B\mid C_N}\le c.
\]
Both $a_W\to1$, retaining the asymptotic minimax result. Moreover the total
variation distance between the two conditioned matrix laws is at most
$(1-\pi_A)^N+(1-\pi_B)^N$, by comparison to their common unconditioned law.
B4's limiting successes are $(0,c)$ and MaxPass's are $(c,0)$; neither is
uniformly superior. The twin's high $\theta'_0$ is strongly disfavored by the
default prior. This standard two-world argument is supporting interpretation,
not an additional originality claim or a contradiction of B4's
prior-approximation guarantee. Exact marginalization has not been shown to
solve this asymptotic family.

\bibliographystyle{ACM-Reference-Format}
\bibliography{references}
\end{document}